\documentstyle[12pt,aps]{revtex}
\begin{document}

\begin{titlepage}
\thispagestyle{empty}
\begin{center}
\vspace*{8mm}
{\Large \bf Time-dependent correlation functions in an one-dimensional
asymmetric exclusion process}\\[22mm]

{\Large {\sc Gunter Sch\mbox{\"u}tz}
} \\[8mm]

\begin{minipage}[t]{13cm}
\begin{center}
{\small\sl
  Department of Physics,
                Weizmann Institute, \\
  Rehovot 76100,
  Israel}
\end{center}
\end{minipage}
\vspace{25mm}
\end{center}
{\small
We continue our studies \cite{1} of an one-dimensional anisotropic
exclusion process with parallel dynamics
describing particles moving to the right
on a chain of $L$ sites. Instead of considering  periodic boundary
conditions with a defect
as in \cite{1} we study  open boundary conditions with injection
of particles
with rate $\alpha$ at the origin and absorption of particles with
rate $\beta$ at the boundary.
We construct the steady state and compute the density profile
as a function of $\alpha$ and $\beta$. In the large
$L$ limit we find a high density phase ($\alpha > \beta$) and a
low density phase ($\alpha < \beta$). In both phases the density
distribution along the chain approaches its respective constant
bulk value exponentially on a length scale $\xi$. They are separated
by a phase transition line where $\xi$
diverges and where the density increases
linearly with the distance from the origin. Furthermore we present
exact expressions for all equal-time n-point density
correlation functions and for the
time-dependent two-point function in the steady state.
We compare our results with predictions from local dynamical scaling
and discuss some conjectures for other exclusion models.
}
\\
\vspace{5mm}\\
\underline{PACS numbers:} 05.40.+j, 05.70.Ln, 64.60.Ht
\end{titlepage}

\newpage
\section{Introduction}
\setcounter{equation}{0}

We study an one-dimensional totally asymmetric exclusion model
where particles are injected stochastically at the origin of a chain
of $L$ sites, move to the right according to rules defined below and
are removed at its end, again according to stochastic rules. Each
site of the chain can be occupied by
at most one particle. Among the interesting features of such
exclusion models is the occurence of various types of phase
transitions which arise from the interplay of the bulk dynamics
with the boundary conditions
\cite{1}-\cite{3} and their close relationship to vertex
models \cite{KDN}, growth models \cite{KS}, and, in the continuum
limit, to the KPZ equation \cite{KPZ} and the noisy Burger's
equation.

Exclusion models can be divided into four classes according to the
dynamics ({\em parallel} or {\em sequential}) and the
boundary conditions
({\em periodic} with conservation of the number of particles and
(possibly) a defect or {\em open} with injection and absorption of
particles). According to this classification we call them
p/p-models (parallel, periodic), p/o-models (parallel, open),
s/p-models (sequential, periodic) and s/o-models (sequential, open).
A s/p-model with
a defect has been studied numerically by Janowsky and Lebowitz
\cite{JL}, without a
defect it was solved by a Bethe ansatz by Gwa and Spohn \cite{GS}.
The s/o-model was studied numerically by Krug \cite{K},
later the exact solution with the full phase diagram was found
\cite{2,3}. In addition to the density profile all equal-time
$n$-point density correlation functions in the steady state were
determined \cite{DE}. Previously we solved a p/p-model with a defect
\cite{1} with Bethe ansatz methods and obtained the density profile
and the equal-time two-point correlation function in the steady state.

Here we discuss the p/o-model with the same parallel dynamics as in
\cite{1} but with  open boundary conditions where particles are
injected at the origin
with a rate $\alpha$ and are removed at the (right) boundary with
rate $\beta$. The bulk dynamics of our model are deterministic and
defined as follows:
Each site $x$ on the ring ($1 \leq x \leq L$) is either occupied
($\tau_x(t) = 1$) or empty ($\tau_x(t) = 0$) at time $t$.
The time evolution consists of two half time steps.
In the first half step we
divide the chain with $L$ sites ($L$ even) into pairs
of sites (2,3), (4,5), $\dots$, $(L,1)$. If both sites in a pair are
occupied or empty or if site $2x$ is empty and site $2x+1$
occupied, they remain so at the intermediate time $t' = t + 1/2$.
If site $2x$ is occupied and
site $2x+1$ empty, then the particle moves with probability 1
to site $2x+1$, i.e.,
\begin{equation}\label{1-1}\begin{array}{lcl}
\tau_{2x}(t') & = & \tau_{2x}(t) \tau_{2x+1}(t) \vspace*{4mm}\\
\tau_{2x+1}(t')   & = & \tau_{2x}(t) + \tau_{2x+1}(t) -
                         \tau_{2x}(t) \tau_{2x+1}(t) \hspace*{2mm} .
\end{array}\end{equation}
These rules are applied in parallel to all pairs except the
pair ($L$,1). In this pair representing the boundary (site $L$) and
the origin (site 1) resp. particles are absorbed and injected
according to the following stochastic rules. If site 1 was empty at
time $t$ then it remains so with probability $1-\alpha$ and becomes
occupied with probability $\alpha$ at time $t'$. If site 1 was
occupied at time $t$ then it remains occupied with probability 1.
These two rules are independent of the occupation of site $L$.
On the other hand, if site $L$ was occupied at time $t$ it remains
so with probability $1-\beta$ and becomes empty with probability
$\beta$. If site $L$ was empty, it remains empty with probability 1.
These two rules are independent of the occupation of site 1.
This means that opposed to the models with sequential dynamics
studied in refs. \cite{K,2,3,DE}
simultaneous injection and absorption is allowed with probability
$\alpha \beta$. We have
\begin{equation}\label{1-2}\begin{array}{rcl}
\tau_1(t') & = & 1 \hspace*{2mm} \mbox{with probability $\tau_1(t) +
                    \alpha (1-\tau_1(t))$} \vspace*{4mm} \\
\tau_1(t') & = & 0 \hspace*{2mm} \mbox{with probability $(1-\alpha)
                           (1-\tau_1(t))$} \vspace*{4mm} \\
\tau_L(t') & = & 1 \hspace*{2mm} \mbox{with probability
$(1-\beta)\tau_L(t)$}
                    \vspace*{4mm} \\
\tau_L(t') & = & 0 \hspace*{2mm} \mbox{with probability $1 - (1-\beta)
                          \tau_L(t)$} \hspace*{2mm} .
\end{array}\end{equation}
In the second half step $t + 1/2 \rightarrow t+1$ the pairing is
shifted by one lattice unit
such that the pairs are now (1,2), (3,4), ... ($L-1,L$). Here rules
(\ref{1-1}) are applied in {\em all} these pairs, there is no
injection and absorption in the second half time step.\footnote{Note
that we reverse the order of the choice of pairs as compared to
\cite{1}. There the pairs were chosen as (1,2), (3,4), ... in
the first half time step.}

In the mapping of ref. \cite{KDN} this model is equivalent to a
two-dimensional four-vertex model in thermal equilibrium with a defect
line where
other vertices, not belonging to the group defining the 6-vertex model
or 8-vertex model, have non-vanishing Boltzmann weights.
The two steps describing the motion of particles
define the diagonal-to-diagonal transfer matrix $T(\alpha,\beta)$
in the vertex model (see appendix). The pairing is chosen
as in \cite{KDN} but the hopping probabilities are different.

For the Bethe ansatz solution of the p/p-model with a defect
\cite{1} the conservation of the number of particles was crucial and
we cannot repeat the calculation here, where the particle number is
not conserved. However, since we are only interested in the
steady state, we can construct the steady state explicitly for
small lattices and then try to guess its general form for arbitrary
length $L$. This method was succesfully applied in refs. \cite{2} and
\cite{3} and led to exact expressions for the particle current
and the density profile for arbitrary values of the injection and
absorption rates. Only after guesswork produced the correct results,
they were actually proven (see also \cite{DE}).
It turns out that also here we can
guess rules for the construction of the steady state. Instead of
proving them we verified our conjecture for lattices of up to 14
sites. In the same way we guessed and verified expressions
for the density profile (the one-point density correlator
\mbox{$\langle \, \tau_x \, \rangle$} in the steady state)
eq. (\ref{2-10}) and the
equal-time $n$-point density correlation function (\ref{4-3}).
The simple form of these correlation functions then
allowed for a conjecture of the time-dependent two-point function
 \mbox{$\langle \, \tau_x T^t \tau_y \, \rangle$} (\ref{5-3}) -
(\ref{5-7}). (In this expression $T^t$ denotes
the $t$-th power of the transfer matrix $T(\alpha,\beta)$).
The mapping to the vertex
model allows for an independent verification of this result.

Among other things the time-dependent two-point function is of
interest for the study of local dynamical scaling in the absence of
translational
invariance. Dynamical scaling in a 1+1 dimensional system with
translational invariance both in space and time direction
implies that the two-point function $G(r,t)$ behaves under a global
rescaling $\lambda$ of the space and time coordinates as \cite{HH}
\begin{equation}\label{1-3}
G(\lambda r,\lambda^z t) = \lambda^{-2x} G(r,t) \hspace*{2mm} .
\end{equation}
In this expression  $r$ denotes the distance in space direction,
$t$ is the distance in time direction, $z$ is the dynamic critical
exponent and $x$ is the scaling dimension. From (\ref{1-3}) follows
that the correlation function has the form
\begin{equation}\label{1-3a}
G(r,t) = t^{-2x/z} \, \Phi(\mbox{\small$\frac{r^z}{t}$})
\end{equation}
with the scaling function $\Phi$ which is not determined by global
dynamical scaling. By extending the concept
of global rescaling to local, space-time dependent rescaling, it has
been shown that for the special case $z=2$ the correlation function
$G(r,t)$ is of the form \cite{Car,Hen}
\begin{equation}\label{1-4}
G(r,t) = a t^{-x} \, \mbox{e}^{-\mbox{\small$\frac{br^2}{2t}$}}
\end{equation}
with some constants $a$ and $b$, i.e., $\Phi(r^2/t)=a\exp{(-br^2/t)}$.
The (connected) density correlation function
in the probabilistic symmetric
p/p-model without defect computed in \cite{KDN} is indeed of this form
with critical exponent $x=1/2$. Since we study the steady state
we have translational invariance in time
direction, but due to the  open boundary conditions translational
invariance is broken
in space direction. In sec.~5 we show that the form of
 \mbox{$\langle \, \tau_x T^t \tau_y \, \rangle$} for large
 $L$ on the
critical line $\alpha=\beta$ and in the scaling regime close to it
resembles (\ref{1-4}) with $x=0$ (sec.~5), i.e., one has $z=2$,
but there are additional pieces that arise from the breaking of
translational invariance.

The paper is organized as follows. In sec. 2 we present our conjectured
rules for the construction of the steady state and exact expressions
for the current and the density profile.
In sec.~3 we study the limit of large $L$ and derive the phase diagram.
In sec.~4 we present expressions for the equal-time $n$-point density
correlation function.
They turn out to be reducible to a sum of one-point functions
through associative fusion rules of the density operators. In
particular we study the two-point function in the scaling regime.
In sec.~5 we compute the time-dependent two-point correlator. Again
we put our emphasis on the vicinity to the phase transition line.
In Sec.~6 we summarize our main results and
discuss our results in the context of other exclusion models. In the
appendix we discuss the mapping to a two-dimensional vertex model.
\section{Construction of the steady state}

Before we discuss the construction of the steady state
we introduce some useful notations.
In anticipation of the correspondence of the model to a vertex model
discussed in refs. \cite{1,KDN} and in the appendix
we denote a state of the system with $N$ particles placed on sites
$x_1, \dots, x_N$ and holes everywhere else
by \mbox{$| \, x_1, \dots, x_N \, \rangle$}. The transfer matrix
$T(\alpha,\beta)$ (\ref{A-1}) acting on the space of states spanned
by these vectors acts as time evolution operator and
encodes the dynamics and the  boundary conditions of the system as
defined by eqs. (\ref{1-1}) and (\ref{1-2}).
The steady state is the (right) eigenvector with eigenvalue 1 of the
transfer matrix $T(\alpha,\beta)$ and we denote it by
\begin{equation}\label{2-1}
\mbox{$| \, 1 \, \rangle$} =
\sum_{N=0}^{L} \sum_{\{ x \} } \Psi_N(x_1 , \dots ,x_N)
          \mbox{$| \, x_1, \dots, x_N \, \rangle$} \hspace*{2mm} .
\end{equation}
Here the $N$-particle ``wave function'' $\Psi_N(x_1, \dots, x_N)$
is the unnormalized probability of finding the particular configuration
\mbox{$| \, x_1, \dots, x_N \, \rangle$} of $N$ particles in the
steady state. We denote
the state with no particles by \mbox{$| \,  \, \rangle$} and the
corresponding wave
function by $\Psi_0$. The
summation runs over all states of $N$ particles ($0\leq N \leq L$)
and all possible configurations $\{x\} = \{ x_1, \dots, x_N \}$
and one has $T(\alpha,\beta) \mbox{$| \, 1 \, \rangle$} =
\mbox{$| \, 1 \, \rangle$}$. The normalized
probabilities are given by
\begin{equation}\label{2-2}
p_N(x_1,\dots,x_N) = \Psi_N(x_1,\dots,x_N)/Z_L
\end{equation}
with
\begin{equation}\label{2-3}
Z_L = \sum_{N=0}^{L} \sum_{\{ x \} } \Psi_N(x_1 , \dots ,x_N)
\hspace*{2mm} .
\end{equation}

The transfer matrix $T(\alpha,\beta)$ has a left eigenvector
\mbox{$\langle \, 1 \, |$}
with eigenvalue 1 given by
\begin{equation}\label{2-3a}
\mbox{$\langle \, 1 \, |$} = \sum_{N=0}^{L} \sum_{\{ x \} }
\mbox{$\langle \, x_1, \dots, x_N \, |$}
\hspace*{2mm} ,
\end{equation}
where \mbox{$\langle \, x_1, \dots, x_N \, |$} is the
transposed vector to
\mbox{$| \, x_1, \dots, x_N \, \rangle$}. Defining a
scalar product in the standard way
(i.e., $\mbox{$\langle \, x_1, \dots, x_N \, |$} y_1,\dots, y_M \,
\rangle = 1$ if the sets
$\{x\}$ and $\{y\}$ are identical and 0 else) one can write $Z_L$
as the scalar product $Z_L = \langle \, 1 \, | \, 1 \, \rangle$.

Furthermore we denote the projection operator on particles on site
$x$ by $\tau_x$:
\begin{equation}\label{2-4}
\tau_x \mbox{$| \, x_1, \dots, x_N \, \rangle$} = \left\{
\begin{array}{ll} \mbox{$| \, x_1, \dots, x_N \, \rangle$}
\hspace*{4mm} &
\mbox{if $x \in \{x_1, \dots, x_N \}$} \vspace*{4mm} \\
0 & \mbox{else} \end{array} \right. \hspace*{2mm} .
\end{equation}
The projector on holes is $\sigma_x = 1-\tau_x$. Expectation values
 \mbox{$\langle \, \tau_{x_1} \dots \tau_{x_k} \, \rangle$}
 of the operators $\tau_x$ and
their products in the steady state can conveniently written in the form
\begin{equation}\label{2-4a}
 \mbox{$\langle \, \tau_{x_1} \dots \tau_{x_k} \, \rangle$} =
\mbox{$\langle \, 1 \, |$} \tau_{x_1} \dots \tau_{x_k}
\mbox{$| \, 1 \, \rangle$} / Z_L \hspace*{2mm}.
\end{equation}
Taking the scalar product with the left eigenvector
\mbox{$\langle \, 1 \, |$} and
dividing by the normalization sum $Z_L$ is equivalent
to a summation over all probabilities $p_N(y_1,\dots,y_N)$ with
$\{x_1,\dots,x_k\} \in \{y_1,\dots,y_N\}$. This is the definition
of an expectation value in the steady state.

The particle current $j$ is a conserved quantity in the bulk since
only the origin and the boundary act as a source or sink of particles.
It is given by the correlator \cite{1} (see (\ref{A-6}))
\begin{equation}\label{2-4b}
j =  \langle \, \tau_{2x}\sigma_{2x+1} \, \rangle
\hspace*{2mm} .
\end{equation}

Now we discuss the construction of the steady state.
In \cite{1} we derived the important result that for the deterministic
dynamics defined by (\ref{1-1}) one has
\begin{equation}\label{2-5}
\tau_{2x-1} \sigma_{2y} \mbox{$| \, \Lambda \, \rangle$} = 0
\end{equation}
for $1\leq x \leq L/2$ and $x \leq y \leq L/2$ and {\em any} right
eigenvector
\mbox{$| \, \Lambda \, \rangle$} of the transfer matrix.
This simplifies the construction
of the steady state considerably: If in a state
\mbox{$| \, x_1, \dots, x_N \, \rangle$} one of
the $x_i$ is odd, then it has a non-vanishing weight
$\Psi_N(x_1,\dots,x_N)$
only if {\em all} even $x_j$ with $x_i < x_j \leq L$ are also contained
in the set $\{ x_1, \dots, x_N \}$.

Using this it is easy to construct the steady state explicitly
for small $L$.
We discovered that the unnormalized probabilities
$\Psi_{N}^{(L+2)}(x_1,\dots,x_N)$ in the chain with $L+2$ sites
can be constructed recursively out those of the chain with $L$ sites
according to the following rules:\vspace*{4mm} \\
\begin{equation}\label{2-6}\begin{array}{ll}
\mbox{\bf \underline{Rule 1:}}
 & (0\leq N \leq L, \mbox{all } \{x\}) \vspace*{4mm} \\
 & \Psi_{N}^{(L+2)}(x_1+2,x_2+2,\dots,x_N+2) = \beta^2(1-\alpha)
\Psi_{N}^{(L)}(x_1,x_2,\dots,x_N) \vspace*{1cm} \\
\mbox{\bf \underline{Rule 2:}}
 & (0\leq N \leq L, \mbox{all } \{x\}) \vspace*{4mm} \\
& \Psi_{N+2}^{(L+2)}(2,x_1,\dots,x_{N-1},L+1,L+2) = \alpha^2(1-\beta)
  \Psi_{N}^{(L)}(2,x_1,\dots,x_{N-1}) \vspace*{1cm} \\
\mbox{\bf \underline{Rule 3:}} & (0\leq N \leq L/2,
\{x_1,\dots,x_{L/2}\} \neq \{2,4,6,\dots,L-2,L\})\vspace*{4mm} \\
 & \Psi_{N+1}^{(L+2)}(2,x_1+2,x_2+2,\dots,x_N+2) = \alpha \beta^2
   \Psi_{N}^{(L)}(x_1,x_2,\dots,x_N) \vspace*{1cm} \\
\mbox{\bf \underline{Rule 4:}} & (L/2 \leq N \leq L,
\{x_1,\dots,x_{L/2}\} \neq \{2,4,6,\dots,L-2,L\})\vspace*{4mm} \\
 & \Psi_{N+1}^{(L+2)}(x_1,x_2,\dots,x_N,L+2) = \alpha^2 \beta
   \Psi_{N}^{(L)}(x_1,x_2,\dots,x_N)  \vspace*{1cm} \\
\mbox{\bf \underline{Rule 5:}} & (\{x_1,\dots,x_{L/2}\} =
\{2,4,6,\dots,L-2,L\}) \vspace*{4mm} \\
 & \Psi_{L/2+1}^{(L+2)}(2,4,\dots,L,L+2) = \alpha \beta (\alpha+\beta)
   \Psi_{L/2}^{(L)}(2,4,\dots,L) - (\alpha\beta)^{L/2+3} \vspace*{1cm}
\end{array}\end{equation}
These rules together with (\ref{2-5}) and the initial conditions
\begin{equation}\label{2-7}
\Psi_0^{(2)} = \beta^2(1-\alpha), \hspace*{4mm}
\Psi_1^{(2)}(2) = \alpha \beta, \hspace*{4mm}
\Psi_2^{(2)}(1,2) = \alpha^2(1-\beta)
\end{equation}
define recursively all quantities
$\Psi_{N}^{(L+2)}(x_1,\dots,x_N)$ in the chain with $L+2$ sites.

Based on these rules we constructed the steady state up to $L=14$
and verified that it has indeed eigenvalue 1 of $T(\alpha,\beta)$.
In a next step we computed the sum over all $\Psi_N(x_1,\dots,x_N)$
and concluded that the normalization $Z_L$ (\ref{2-3}) is given by
\begin{equation}\label{2-8}\begin{array}{rcr}
Z_L & = & \displaystyle (1-\beta) \alpha^L
          \sum_{k=0}^{L/2-1} \left( \frac{\beta}{\alpha} \right)^k +
          (1-\alpha) \beta^L
          \sum_{k=0}^{L/2-1} \left( \frac{\alpha}{\beta} \right)^k +
          (\alpha \beta)^{L/2} \vspace*{4mm} \\
    & = & \left\{ \begin{array}{ll} \displaystyle
                 \frac{(1-\beta)\alpha^{L+1} - (1-\alpha)\beta^{L+1}}
    {\alpha-\beta} \hspace*{1cm} & \alpha \neq \beta \vspace*{4mm} \\
    \alpha^L \left( 1+L(1-\alpha) \right) & \alpha=\beta\end{array}
          \right. \hspace*{2mm} .
\end{array}\end{equation}
This result was again checked explicitly up to $L=14$.

Going one step further we consider the average density
\mbox{$\langle \, \tau_x \, \rangle$}
at site $x$ defined by (\ref{2-4a}). We found the following exact
expressions for the even and odd sublattices resp.:
\begin{equation}\label{2-9}\begin{array}{lcl}
Z_L  \mbox{$\langle \, \tau_{2x} \, \rangle$} & = &
          \left\{ \begin{array}{l} \displaystyle  (1-\beta) \alpha^L
          \sum_{k=0}^{2x-1} \left( \frac{\beta}{\alpha} \right)^k +
          (1-\beta) \alpha^{L+1}
          \sum_{k=2x}^{L} \left( \frac{\beta}{\alpha} \right)^k +
          (\alpha \beta)^{L+1} \vspace*{4mm} \\
          \alpha^{L+1} \left( 1+L(1-\alpha) \right) +
          2x \alpha^L (1-\alpha)^2  \end{array} \right. \vspace*{6mm}\\
Z_L  \mbox{$\langle \, \tau_{2x-1} \, \rangle$} & = &
          \left\{ \begin{array}{l} \displaystyle  (1-\beta)^2 \alpha^L
          \sum_{k=0}^{2x-3} \left( \frac{\beta}{\alpha} \right)^k +
          (1-\beta) \alpha^{L+2-2x} \beta^{2x-2} \vspace*{4mm} \\
          \alpha^{L+1} (1-\alpha) + (2x-1)
          \alpha^L (1-\alpha)^2  \end{array} \right. \hspace*{2mm} .
\end{array}\end{equation}
On the r.h.s. of (\ref{2-9}) the upper expressions are valid for
$\alpha \neq \beta$, while the lower expressions are valid for
$\alpha = \beta$. Performing the summations we arrive at the main
result of this section
\begin{equation}\label{2-10}\begin{array}{lcl}
 \mbox{$\langle \, \tau_{2x} \, \rangle$} & = &
 \left\{ \begin{array}{ll} \displaystyle
\alpha + (1-\alpha) \frac{1-(\mbox{\small$\frac{\beta}{\alpha})^{2x}$}}
{1- \mbox{\small$\frac{1-\alpha}{1-\beta}$}
(\mbox{\small$\frac{\beta}{\alpha}$})^{L+1}} &
\alpha \neq \beta \vspace*{4mm} \\
\displaystyle \alpha + (1-\alpha)^2 \frac{2x}{1+L(1-\alpha)} &
\alpha = \beta \end{array} \right.
\vspace*{6mm} \\
 \mbox{$\langle \, \tau_{2x-1} \, \rangle$} & = &
 \left\{ \begin{array}{ll} \displaystyle (1-\beta)
\frac{1-\mbox{\small$\frac{1-\alpha}{1-\beta}$}
(\mbox{\small$\frac{\beta}{\alpha}$})^{2x-1}}
{1- \mbox{\small$\frac{1-\alpha}{1-\beta}$}
(\mbox{\small$\frac{\beta}{\alpha}$})^{L+1}} &
\alpha \neq \beta \vspace*{4mm} \\
\displaystyle (1-\alpha)^2 \frac{2x}{1+L(1-\alpha)} +
\frac{\alpha(1-\alpha)}{1+L(1-\alpha)} &
\alpha = \beta \end{array} \right. \hspace*{2mm} .
\end{array}\end{equation}

The anisotropy between the even and odd sublattices is a consequence
of the parallel updating mechanism \cite{1} and related to the
net particle current (\ref{2-4b}) for which we found
(again, by direct evaluation on small lattices and guessing the
general form for arbitrary $L$)
\begin{equation}\label{2-11}
j = \left\{ \begin{array}{ll} \displaystyle \alpha -
\frac{\alpha - \beta}{1- \mbox{\small$\frac{1-\alpha}{1-\beta}$}
(\mbox{\small$\frac{\beta}{\alpha}$})^{L+1}} &
\alpha \neq \beta \vspace*{4mm} \\
                    \displaystyle \alpha -
\frac{\alpha(1-\alpha)}{1+L(1-\alpha)} &
\alpha = \beta \end{array} \right. \hspace*{2mm} .
\end{equation}
The fact that this quantitity is independent on $x$ as it should
be is an additional non-trivial check for our conjectures.

Finally note that by the definition of the model there is the
particle-hole symmetry (\ref{A-5}): Changing particles
into holes, reflecting
site $x$ into site $L+1-x$ and exchanging $\alpha$ and $\beta$
leaves the system invariant. All our results are indeed invariant
under this operation.

\section{The phase diagram}

{}From eqs. (\ref{2-10}) one realizes that the system changes its
behaviour if $\alpha=\beta$. The average densities on the even
and odd sublattices
\begin{equation}\label{3-1}
\rho^{(even)} = \frac{2}{L} \sum_{x=1}^{L/2}
\mbox{$\langle \, \tau_{2x} \, \rangle$}, \hspace*{4mm}
\rho^{(odd)} = \frac{2}{L} \sum_{x=1}^{L/2}
\mbox{$\langle \, \tau_{2x-1} \, \rangle$}
\end{equation}
have a discontinuity in the thermodynamic limit $L \rightarrow \infty$
at $\alpha=\beta\neq 0,1$.\footnote{If $\alpha$ or $\beta$ is
either 0 or 1 the system is trivial in the sense that the
density profile is exactly constant (even in finite systems)
and all correlation functions can be obtained without any
calculation (see appendix). Therefore we exclude these cases
from our discussion.} One finds from (\ref{2-10})
\begin{equation}\label{3-2}\begin{array}{rcr}
\rho^{(even)} & = &
            \left\{ \begin{array}{ll} \alpha  \hspace*{1cm} &
            \alpha < \beta \vspace*{4mm} \\
                         1         & \alpha > \beta \end{array} \right.
            \vspace*{6mm} \\
\rho^{(odd)} & = &
              \left\{ \begin{array}{ll} 0 \hspace*{1cm} &
              \alpha < \beta \vspace*{4mm} \\
                            1 - \beta & \alpha > \beta
                            \end{array} \right.
\end{array}\end{equation}
On the phase transition line $\alpha=\beta$ one obtains
$\rho^{(even)} = (1+\alpha)/2$ and $\rho^{(odd)} =(1-\alpha)/2$ resp.
For $\alpha < \beta$ (more particles are absorped than injected)
the system is in a low density phase with total average density
$\rho = \mbox{\small$\frac{1}{2}$}, \,
(\rho^{(even)} + \rho^{(odd)}) = \alpha/2 < \mbox{\small$\frac{1}{2}$}$,
while for $\alpha > \beta$ it is in a high density phase with
$\rho = 1-\beta/2 > \mbox{\small$\frac{1}{2}$}$ (Fig.~1).

\begin{figure}
\setlength{\unitlength}{1.5mm}
\begin{center}
\begin{picture}(50,50)
\put (4,4){0}
\put (1,25){1/2}
\put (3,45){1}
\put (25,3){1/2}
\put (45,3){1}
\put (26,0){$\alpha$}
\put (-2,25){$\beta$}
\put (6,6){\framebox(40,40)}
\put (17,32){A}
\put (32,17){B}
\put (6,6){\line(1,1){40}}
\end{picture}
\end{center}
\caption{\protect\small
Phase diagram of the model in the $\alpha-\beta$ plane.
Region A is the low density phase and region B  the high density phase.
The phases are separated by the curve $\alpha=\beta$.}
\end{figure}
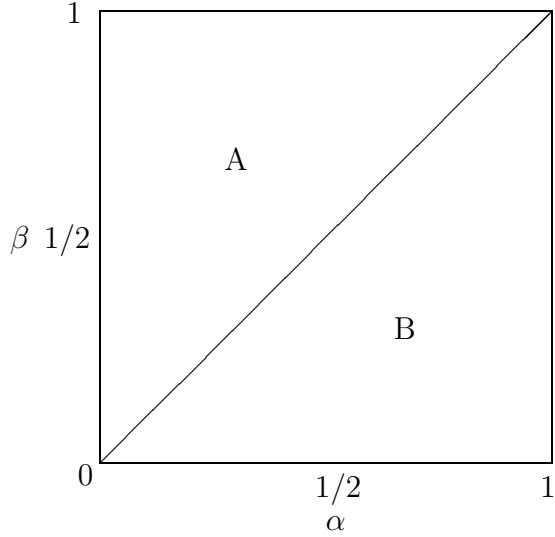

In the thermodynamic limit $L \rightarrow \infty$
the current $j$ (\ref{2-11}) is given by
\begin{equation}\label{3-3}
j = \min{(\alpha,\beta)} \hspace*{2mm} .
\end{equation}
There is no discontinuity at $\alpha=\beta$ in the current, but its
first derivatives w.r.t. $\alpha$ and $\beta$ are discontinous. The
discontinuity of $\rho$ at the phase transition line $\alpha=\beta$
and eq. (\ref{3-3}) remind us of the s/o-model \cite{2,3,DE}.
In this model in the region $\alpha,\beta < 1/2$
the phase diagram shows a low density phase ${\rm A}_{\rm I}$
and a high density phase ${\rm B}_{\rm I}$
separated by a phase transition line at $\alpha=\beta$ \cite{3}.
Also in this model the density and the first derivatives of the
current w.r.t. the injection and absorption rates $\alpha$ and $\beta$
have a discontinuity at the phase transition line.

In terms of the sublattice densities the current $j$ is given by $j=
\rho^{(even)} - \rho^{(odd)}$ for all $\alpha,\beta$. In terms
of the total average density $\rho$ the current satisfies
$j=2\rho$ if $\rho<1/2$ (low density phase) and $j=2(1-\rho)$
if $\rho>1/2$ (high density phase). These are the same relations
as in the p/p-model in the respective phases \cite{1}.

Now we turn to a discussion of the density profile.
We first study the case $\alpha<\beta$ and $L \rightarrow \infty$.
Defining the decay length $\xi$ by
\begin{equation}\label{3-4}
\xi^{-1} = \ln{\frac{\beta}{\alpha}}
\end{equation}
one obtains from (\ref{2-10}) the density profile up to corrections
of order $\exp{(-L/\xi)}$
\begin{equation}\label{3-5}\begin{array}{rcl}
 \mbox{$\langle \, \tau_{2x} \, \rangle$} & = & \alpha +
 (1-\beta) \mbox{e}^{-(L+1-2x)/\xi}
                        \vspace*{4mm} \\
 \mbox{$\langle \, \tau_{2x-1} \, \rangle$} & = &
 (1-\beta) \mbox{e}^{-(L+2-2x)/\xi} \hspace*{2mm} .
\end{array}\end{equation}
The profile decays exponentially with increasing distance from the
boundary to its respective bulk values $\rho_{bulk}^{(even)} =
\alpha$ and $\rho_{bulk}^{(odd)} = 0$. This the low density phase
of the system.

In the high density phase $\alpha > \beta$ which is related to
the low density phase by the particle-hole symmetry the profile
is given by
\begin{equation}\label{3-6}\begin{array}{rcl}
 \mbox{$\langle \, \tau_{2x} \, \rangle$} & = & 1 -
 (1-\alpha)\mbox{e}^{-2x/\xi}
                        \vspace*{4mm} \\
 \mbox{$\langle \, \tau_{2x-1} \, \rangle$} & = & 1-\beta -
 (1-\alpha)\mbox{e}^{-(2x-1)/\xi}
\hspace*{2mm} .
\end{array}\end{equation}
The bulk densities are $\rho_{bulk}^{(even)} = 1$ and
$\rho_{bulk}^{(odd)} = 1-\beta$.

On approaching the phase transition line $\alpha=\beta$ the decay
length $\xi$ diverges. On the line the profile is linear
and up to corrections of order $L^{-1}$ given by
\begin{equation}\label{3-7}\begin{array}{rcl}
 \mbox{$\langle \, \tau_{2x} \, \rangle$} & = & \displaystyle
 \alpha + (1-\alpha)\frac{2x}{L}\vspace*{4mm}\\
 \mbox{$\langle \, \tau_{2x-1} \, \rangle$} & = & \displaystyle
 (1-\alpha)\frac{2x-1}{L} \hspace*{2mm} .
\end{array}\end{equation}
An explanation for the shape of the profile in the two phases and
on the phase transition line will be given in the next section.

\section{Equal-time correlation functions}

Having found exact expressions for the current and the density profile
we proceed calculating the $n$-point equal-time density
correlation function  \mbox{$\langle \,
\tau_{x_1} \dots \tau_{x_n} \, \rangle$}
in the steady state.
Examining the two-point function for small $L$ we found the following
{\em exact} relations for $1\leq x < y \leq L/2$
\begin{equation}\label{4-1}\begin{array}{rcl}
 \mbox{$\langle \, \tau_{2x} \tau_{2y} \, \rangle$}& = &
 \mbox{$\langle \, \tau_{2x} \, \rangle$} - \alpha +
 \alpha  \mbox{$\langle \, \tau_{2y} \, \rangle$} \vspace*{4mm} \\
 \mbox{$\langle \, \tau_{2x} \tau_{2y-1} \, \rangle$}& = &
(1-\beta)( \mbox{$\langle \, \tau_{2x} \, \rangle$}-\alpha)+
\alpha  \mbox{$\langle \, \tau_{2y-1} \, \rangle$} \vspace*{4mm} \\
 \mbox{$\langle \, \tau_{2x+1} \tau_{2y} \, \rangle$}& = &
 \mbox{$\langle \, \tau_{2x+1} \, \rangle$} \vspace*{4mm} \\
 \mbox{$\langle \, \tau_{2x-1} \tau_{2y-1} \, \rangle$}& = &
(1-\beta) \mbox{$\langle \, \tau_{2x-1} \, \rangle$}\hspace*{2mm} .
\end{array}\end{equation}
The third of these equations is a simple consequence of (\ref{2-5})
which says that whenever there is a particle on an odd lattice site
then all even lattice sites to its right must be occupied as well.
The important result is that
the two-point is completely determined by the one-point function and
some constants! Going further we made the surprising observation
that the $n$-point function can also be expressed in terms of one-point
functions by repeatedly fusing products of {\em operators}
$\tau_x \tau_y$ according to the fusing rules that are defined by
(\ref{4-1}) by omitting the averaging. This fusion can be performed
in arbitrary order until one reaches the one-point level.

The fusion rules implied by eqs. (\ref{4-1}) can be simplified by using
operators $\eta_x$ defined by
\begin{equation}\label{4-1a}
\eta_{2x} = \frac{\tau_{2x} - \alpha}{1-\alpha}, \hspace*{1cm}
\eta_{2x-1} = \frac{\tau_{2x-1}}{1-\beta}
\end{equation}
instead of using the density operators $\tau_x$. In the bulk of the
high density region both  \mbox{$\langle \, \eta_{2x} \, \rangle$} and
 \mbox{$\langle \, \eta_{2x-1} \, \rangle$} take the
value 1, while in the bulk of the low density region both average
values are 0.
Expressing all $\tau_x$ in terms of the $\eta_x$ the correlation
functions (\ref{4-1}) become
\begin{equation}\label{4-2}
 \mbox{$\langle \, \eta_{x_1}\eta_{x_2} \, \rangle$} =
 \mbox{$\langle \, \eta_{x_1} \, \rangle$}\hspace*{2cm}
 (x_2>x_1)\hspace*{2mm} .
\end{equation}
Fusion of $n$ operators $\eta_{x_1}\dots\eta_{x_n}$ gives $\eta_{x_i}$
with $x_i = \min{\{x_1,\dots,x_n\}}$. So the $n$-point correlation
function is
\begin{equation}\label{4-3}
 \mbox{$\langle \, \eta_{x_1}\dots \eta_{x_n} \, \rangle$}
 = \mbox{$\langle \, \eta_{x_i} \, \rangle$} \hspace*{2cm}
(x_i = \min{\{x_1,\dots,x_n\}}) \hspace*{2mm} .
\end{equation}
This is the main result of this section.

The form of the two-point function (\ref{4-2}) can be understood by
considering the steady state  as composed of ``constituent profiles''
with a region of constant low density up to some point $x_0$
in the chain
followed by a high density region beyond this ``domain wall''.
Such an assumption
explains why the correlator (\ref{4-2}) does not depend on $x_2$:
In the low density region of density $\alpha$ on the even sublattice
and 0 on the odd sublattice the operator $\eta_{x_1}$ has vanishing
expectation value
and therefore the whole expression
\mbox{$\langle \, \eta_{x_1}\eta_{x_2} \, \rangle$}
is zero if $x_1$ is in this region, independent of $\eta_{x_2}$.
If, however, $x_1$ is in
a region of high density, then, according to our assumption, also
$x_2>x_1$ must be in region of high density. Thus,
$\eta_{x_1}\eta_{x_2}$ again does not depend on $x_2$ and takes
the value 1. We conclude that the product $\eta_{x_1}\eta_{x_2}$
is either 0 or 1, depending on whether $x_1$ is in a region of
low or high density. This leads to the expression (\ref{4-2}) for
the expectation value of this product.

The average value  \mbox{$\langle \, \eta_x \, \rangle$}
itself contains the information about the position
$x_0$ of the domain wall. In the low density phase the density profile
decays exponentially from above to its bulk value with increasing
distance from the boundary. This means that the probability of finding
the domain wall also decreases exponentially with the same decay
length $\xi$ with the distance from the boundary. The domain wall
is caused by particles hitting the boundary where they get stuck
with probability $1-\beta$ and then cause other incoming particles
to pile up and create a region of high density (fig.~4 in the
appendix). On the other hand,
in the high density phase where the rate of injection is higher than
the absorption rate
the situation is reversed. Here the probability
of finding the domain wall decreases exponentially
with the distance from the origin. This can alternatively be explained
either in terms of holes through the particle-hole symmetry or in terms
of particles being piled up from the boundary over the whole system
up to a point close to the origin. On the phase transition line
injection and absorption are in balance and the probability of
finding the domain wall is space-independent and the density
profile is a linear superposition of the assumed step function
constituent profiles. This leads to the observed linearly increasing
average density (\ref{2-7}).

We conclude this section by studying the two-point function in more
detail in the limit $L\rightarrow \infty$.
We define the equal-time connected two-point function by
\begin{equation}\label{4-4}
G_c(x_1,x_2;t=0) =
 \mbox{$\langle \, \eta_{x_1}\eta_{x_2} \, \rangle$} -
 \mbox{$\langle \, \eta_{x_1} \, \rangle$}
 \mbox{$\langle \, \eta_{x_2} \, \rangle$} =
\frac{ \mbox{$\langle \, \tau_{x_1}\tau_{x_2} \, \rangle$} -
\mbox{$\langle \, \tau_{x_1} \, \rangle$}
 \mbox{$\langle \, \tau_{x_2} \, \rangle$}}
 {(1-\alpha)^m(1-\beta)^n}  \hspace*{2mm} .
\end{equation}
where $m=2,n=0$ if both $x_1$ and $x_2$ are even,
$n=2,m=0$ if both $x_1$ and $x_2$ are odd and $m=n=1$ else.

In what follows we restrict ourselves to the case where $x_1$ and
$x_2$ are both odd, the mixed correlators can be computed analogously.
{}From (\ref{4-2}) one obtains $G(x_1,x_2;0)
=  \mbox{$\langle \, \eta_{x_1} \, \rangle$}
(1- \mbox{$\langle \, \eta_{x_2} \, \rangle$})$
and inserting the expressions for
the density profile (\ref{2-5}) - (\ref{2-7}) one obtains with
$x_2 = x_1 + 2r$ ($r>0$)
\begin{equation}\label{4-5}
G_c(x_1,x_2;0) = \left\{ \begin{array}{ll} \displaystyle
A(x_2) \mbox{e}^{-2r/\xi} \hspace*{1cm} & \alpha < \beta
\vspace*{4mm} \\
\tilde{A}(x_1)\mbox{e}^{-2r/\xi}
\hspace*{1cm} & \alpha > \beta \vspace*{4mm} \\
\frac{x_1}{L} (1-\frac{x_1}{L}) - \frac{x_1}{L}\frac{r}{L} &
\alpha=\beta
\end{array} \right.
\hspace*{2mm} .
\end{equation}
The amplitudes of the exponential decay are given by
\begin{equation}\label{4-6}\begin{array}{rcl}
A(x) & = & \displaystyle \mbox{e}^{-R/\xi}
\left( 1-\mbox{e}^{-R/\xi} \right) \vspace*{4mm} \\
\tilde{A}(x) & = & \displaystyle \frac{1-\alpha}{1-\beta}
\mbox{e}^{-x/\xi}
\left( 1-\frac{1-\alpha}{1-\beta}\mbox{e}^{-x/\xi} \right)
\end{array}\end{equation}
where $R=L+1-x$ measures the distance of site $x$ from the boundary.

The decay length $\xi$ is identical with correlation length of the
connected two-point function. On the phase transition line the
correlation function is constant for relative distances $2r \ll L$.
Its amplitude depends on the position $x_1$ in the bulk. A similar
form of the connected equal-time correlator was found in the
p/p-model with a defect \cite{1}.

\section{Time-dependent correlation functions}

In this section we study the time-dependent two-point correlation
function in the steady state
\begin{equation}\label{5-1}
G(x_1,x_2;t) =  \mbox{$\langle \, \eta_{x_1}T^t\eta_{x_2} \, \rangle$}
\end{equation}
where $T^t$ denotes the $t$-th power of the transfer matrix
$T$.\footnote{Note that $t=1$ corresponds to a distance of two lattice
units in time direction in the underlying vertex model
(see appendix).} We define the direction of the time evolution
formally by $T^{-t}\tau_x(t_0)T^t= \tau_x(t_0+t)$, thus
$G(x_1,x_2;t) =  \mbox{$\langle \, \eta_{x_1}(t_0+t)\eta_{x_2}(t_0)
\, \rangle$}$.
The connected two-point function is defined by
$G_c(x_1,x_2;t)=G(x_1,x_2;t) - \mbox{$\langle \, \eta_{x_1} \, \rangle$}
 \mbox{$\langle \, \eta_{x_2} \, \rangle$}$.

The standard way of computing the correlation function (\ref{5-1})
would be the insertion of a complete set of eigenstates of $T$,
evaluating the matrix elements
$a_k(x_1)=\mbox{$\langle \, 1 \, |$}\eta_{x_1}\mbox{$| \,
\Lambda_k \, \rangle$}$ and
$\tilde{a}_k(x_2)=\mbox{$\langle \, \Lambda_k \, |$}\eta_{x_2}
\mbox{$| \, 1 \, \rangle$}$ and
summing over $a_k\tilde{a}_k\Lambda_k^t$.
Since we do not know the eigenstates and eigenvalues we take the
alternative route using the  commutator of
$[\, \eta_x, \, T^t \, ]$.
Since \mbox{$\langle \, 1 \, |$} is a left eigenvector of $T$ with
eigenvalue 1 one
has $ \mbox{$\langle \, [\, \eta_x, \, T^t \, ]
\eta_{x_2} \, \rangle$}=
\mbox{$\langle \, 1 \, |$}(\eta_{x_1}T^t
- T^t \eta_{x_1})\eta_{x_2}\mbox{$| \, 1 \, \rangle$} = G_c(x_1,x_2;t)$.
{}From this one obtains $G(x_1,x_2;t)$.

First we note that
from the commutation relations (\ref{A-7}) of $\tau_x$ with $T$
one obtains
\begin{equation}\label{5-1a}\begin{array}{rcl}
\tau_{2x-1} \, T & = & T \, (1 - \sigma_{2x-2}\sigma_{2x-1})
                      \tau_{2x}\tau_{2x+1} \vspace*{4mm} \\
\tau_{2x} \, T   & = & T \, (1 -
\sigma_{2x-2}\sigma_{2x-1}(1-\tau_{2x}\tau_{2x+1})) \hspace*{2mm} .
\end{array}\end{equation}
It is obvious that evaluating $\tau_x\,T^t$
is not an easy task. By iterating relations (\ref{5-1a})
$t$ times not only the number of terms in the products on the r.h.s.
but also the total number of such multi-point correlators increases
extremely fast with $t$. It is only the simplicity of the multi-point
correlators (see eqs. (\ref{4-2}) and (\ref{4-3})) that makes this
approach promising. We restrict our discussion again to both
$x_1=2y_1-1$ and $x_2=2y_2-1$ odd.

By iterating (\ref{5-1a}) $t$ times one finds that
$\tau_{2y_1-1} T^t\tau_{2y_2-1}$ is of the form
\begin{equation}\label{5-2}\begin{array}{rcl}
\tau_{2y_1-1} T^t \tau_{2y_2-1} & = & \displaystyle T^t \left\{
1-(\sigma_{2y_1-2t}\sigma_{2y_1-2t+1}\dots\hspace*{4mm}) -
(\sigma_{2y_1-2t+2}\sigma_{2y_1-2t+3}\dots\hspace*{4mm}) - \right.
\vspace*{4mm} \\
 & & \hspace*{4mm} (\hspace*{4mm}\dots\hspace*{4mm}) -
 (\sigma_{2y_1-2t+2k}\sigma_{2y_1-2t+2k+1}\dots\hspace*{4mm})
  - (\hspace*{4mm}\dots\hspace*{4mm}) - \vspace*{4mm} \\
 & & \hspace*{4mm} \left. \sigma_{2y_1-4+2t} \sigma_{2y_1-3+2t}
 \right\} \tau_{2y_1-2+2t}\tau_{2y_1-1+2t}\tau_{2y_2-1}
\end{array}\end{equation}
where the dots denote some complicated sums of products of operators
$\tau_y$ acting on sites $y$ between
$2y_1-2t$ and $2y_1-1+2t$. $\sigma_y =1-\tau_y$ denotes
the projector on holes and in order to avoid boundary effects one
has to choose $t<y_1 - 1$. We first discuss the correlation function
outside the light cone, then on the edges of the light cone and
finally in its interior.

\subsection{Correlation function outside the light cone}

We want to evaluate  \mbox{$\langle \, \tau_{2y_1-1}T^t\tau_{2y_2-1}
\, \rangle$} with
$2y_1-1 \geq 2y_2-1+2t$. Recalling the fusion rule
$ \mbox{$\langle \, \tau_{2x+1}\tau_{2y} \, \rangle$}
= \mbox{$\langle \, \tau_{2x+1} \, \rangle$}$ for
$y>x$ (\ref{4-1}) one obtains $ \mbox{$\langle \, \tau_{2x+1}\sigma_{2y}
\, \rangle$}=0$ for $y>x$.
Since the fusion procedure is associative all terms on the r.h.s. of
(\ref{5-2}) vanish when contracted with $\tau_{2y_2-1}$ except
$ \mbox{$\langle \, \tau_{2y_2-1}\tau_{2y_1-2+2t}\tau_{2y_1-1+2t}
\, \rangle$}=
 \mbox{$\langle \, \tau_{2y_1-1-2t}\tau_{2y_1-2+2t}\tau_{2y_1-1+2t}
\, rangle$ }$.
Using also $ \mbox{$\langle \, \tau_{2x-1}\tau_{2y-1} \, \rangle$}
=(1-\beta)  \mbox{$\langle \, \tau_{2x-1} \, \rangle$}$ for $y>x$
one gets
$ \mbox{$\langle \, \tau_{2y_1-1-2t}\tau_{2y_1-2+2t}
\tau_{2y_1-1+2t} \, \rangle$}=(1-\beta)
 \mbox{$\langle \, \tau_{2y_1-1-2t} \, \rangle$}$. Therefore we obtain
\begin{equation}\label{5-3}
G(x_1,x_2;t) =  \mbox{$\langle \, \eta_{x_2} \, \rangle$} \hspace*{2cm}
(\mbox{$x_1,x_2$ odd, $x_1 \geq x_2 +2t$})\hspace*{2mm} .
\end{equation}

Now we study the correlator
\mbox{$\langle \, \tau_{2y_1-1}T^t\tau_{2y_2-1} \, \rangle$} with
$2y_1-1 \leq 2y_2-3-2t$. Here the fusion of $\tau_{2y_1-1+2t}$
in the r.h.s. of (\ref{5-2})
with $\tau_{2y_2-1}$ yields $(1-\beta)\tau_{2y_1-1+2t}$ and by taking
the average value one obtains $ \mbox{$\langle \,
\tau_{2y_1-1}T^t\tau_{2y_2-1} \, \rangle$} =
(1-\beta) \mbox{$\langle \, \tau_{2y_1-1}T^t \,
\rangle$} = (1-\beta) \mbox{$\langle \, \tau_{2y_1-1} \, \rangle$}$.
We find
\begin{equation}\label{5-4}
G(x_1,x_2;t) =\mbox{$\langle \, \eta_{x_1} \, \rangle$} \hspace*{2cm}
(\mbox{$x_1,x_2$ odd, $x_1 \leq x_2 -2-2t$})\hspace*{2mm} .
\end{equation}

Eqs. (\ref{5-3}) and (\ref{5-4}) are no surprise.
The area defined by (\ref{5-3}) and (\ref{5-4}) is
the exterior of the forward light cone of the particle at site $x_2$.
If $x_1$ and $x_2$  are chosen in this way and both are in a region of
uniform density (either in the bulk of the high density phase or in
the bulk of the low density phase) one has $\eta_{x_1}=\eta_{x_2}=1
\mbox{ or } 0$ and therefore
the connected correlation function $G_c(x_1,x_2;t)$
is time-independent and 0 as one would expect.
In the boundary region of the low density phase where particles
pile up and lead to a non-uniform density profile
(or near the origin
in the high density phase) it is still time-independent as it must
be outside the light cone, but non-zero (see (\ref{4-5})).
This is due to the hard-core repulsion
of the particles which behave as an incompressible liquid.

\subsection{Correlation function on the edges of the light cone}

On the right edge of the light cone of the particle at
site $x_2$ defined by
$x_1 = x_2 - 2 +2t$ we can repeat the considerations that led to
(\ref{5-3}): All the pieces on the r.h.s. of (\ref{5-2}) containing
$\tau_{2y_2-1}\sigma_{2y_1-2t+2k}=
\tau_{2y_1+1-2t}\sigma_{2y_1-2t+2k}$
with $k\geq 1$ vanish as a result of the fusion rules and only
the first two pieces in the sum remain. Although the term containing
$\sigma_{2y_1-2t}\sigma_{2y_1+1-2t}\dots\tau_{2y_1+1-2t}$ does not
vanish due to fusion
with $\tau_{2y_1+1-2t}$ it is nevertheless 0 since by definition
$\sigma_{2y_1+1-2t}\tau_{2y_1+1-2t}=0$. Therefore
\begin{equation}\label{5-5}
G(x_2-2+2t,x_2;t) =  \mbox{$\langle \, \eta_{x_2} \, \rangle$}
\hspace*{2cm} (\mbox{$x_2$ odd}) \hspace*{2mm}.
\end{equation}
Consequently the connected correlation function on the odd sublattice
vanishes also on the forward edge of the light cone if the two points
are in a region of uniform density.
This is a result of the asymmetry of the model: if the system is in a
region of uniform low density $\rho<1/2$
the odd sublattice is empty and the vanishing of the
correlator is trivial. In a region of uniform high density
the even sublattice
is completely occupied and particles on the odd sublattice effectively
move only to the left (fig.~4 in the appendix) and are therefore
uncorrelated to particles on the right edge of their (forward)
light cone.

Due to the deterministic nature of the dynamics the particles on the
odd sublattice move with the velocity of light, i.e.
two lattice units per full time step as long as they are in region of
uniform high density. Thus we expect a singularity of the correlation
function on the left edge of the light cone defined by $x_1 = x_2 -2t$:
Indeed, choosing $2y_2-1=2y_1-1+2t$ does not change the r.h.s.
of (\ref{5-2}) since $\tau_{2y_1-1+2t}^2=\tau_{2y_1-1+2t}$
and therefore $ \mbox{$\langle \,
\tau_{2y_1-1}T^t\tau_{2y_1-1+2t} \, \rangle$} =
 \mbox{$\langle \, \tau_{2y_1-1}T^t \, \rangle$} =
 \mbox{$\langle \, \tau_{2y_1-1} \, \rangle$}$.
For the correlator (\ref{5-1}) we obtain
\begin{equation}\label{5-6}
G(x_2-2t,x_2;t) =  (1-\beta)^{-1}
 \mbox{$\langle \, \eta_{x_2-2t} \, \rangle$} \hspace*{2cm}
 (\mbox{$x_2$ odd}) \hspace*{2mm} .
\end{equation}
Here the connected correlation function in a region of uniform
high density does not vanish.

\subsection{Correlation function inside the light cone}

First we note that in a region of uniform high density the result
is again trivial. In such a region particles on the odd sublattice
are found everywhere with equal probability (the equal-time connected
two-point function is 0) and since they move with the velocity of
light the time-dependent connected two-point function does also
vanish.

If the profile is not uniform the calculation inside the
light cone is non-trivial. With $x_1=2y_1-1$
as above and $x_2$ increasing beyond $2y_1+2-2t$ more and more
contributions from the r.h.s. of (\ref{5-2}) are non-zero.
We evaluated $G(x_1,x_2;t)$ for $t=1,2,3$ inside the light cone
on the computer (using the software system {\em Mathematica}
\cite{mat}) by calculating the exact form of (\ref{5-2})
and then implementing the fusion rules (\ref{4-1}) on
the multi-point correlators on the r.h.s. of (\ref{5-2}).
First we noticed that in a region of uniform high (low) density
(all  \mbox{$\langle \, \eta_x \, \rangle$}=1(0)) one obtains
$G(x_1,x_2;t)=1(0)$ and
therefore $G_c(x_1,x_2;t)=0$ as it should be.
This observation is indeed a highly non-trivial test
of the conjectured fusion rules (\ref{4-1}) on which our
calculation is based: Since eqs. (\ref{4-1}) are supposed to be
exact, the result
of the calculation of the time-dependent correlator must also
be exactly 1 (0) if all $\eta_x$ involved are set to 1 (0).
Any other result would have shown that the fusion rules do not hold.
Secondly we observed that by taking $\alpha=1-\beta$ the exact general
form of the correlator becomes fairly obvious for arbitrary values
of $x_1,x_2$ and $t$ with $x_1$ inside the light cone of $x_2$.
We found by generalizing our result from $t=1,2,3$ to arbitrary $t$
\begin{equation}\label{5-7}\begin{array}{rcl}
G(x,x+2y;t) & = & \displaystyle \sum_{k=0}^{t+y-2} \left\{
{\renewcommand{\arraystretch}{0.6}
\mbox{$\left(\begin{array}{@{}c@{}}{\scriptstyle 2t-2}\\
{\scriptstyle k}\end{array}\right)$}
\renewcommand{\arraystretch}{1}}
\beta^{2t-2-k}(1-\beta)^k \cdot \right.  \vspace*{4mm} \\
 & & \displaystyle \left. \hspace*{1cm}
 \left( \beta \mbox{$\langle \, \eta_{x+3-2t+2k} \, \rangle$} +
(1-\beta) \mbox{$\langle \, \eta_{x+4-2t+2k} \,
\rangle$}\right)\right\} +  \vspace*{4mm} \\
 & & \displaystyle \left( 1 - \sum_{k=0}^{t+y-2}
{\renewcommand{\arraystretch}{0.6}
\mbox{$\left(\begin{array}{@{}c@{}}{\scriptstyle 2t-2}\\
{\scriptstyle k}\end{array}\right)$}
\renewcommand{\arraystretch}{1}}
\beta^{2t-2-k}(1-\beta)^k \right)  \mbox{$\langle \,
\eta_{x+2y} \, \rangle$} \hspace*{2mm} .
\end{array}\end{equation}
This the main result of this section, valid for $t<(x-1)/2$
and $-t + 2 \leq y \leq t -1$. The first restriction is due to
boundary effects, the second defines the interior of the light cone.
In order to check this result we explicitly calculated
$G(x,x+2y;t=4)$ on the computer for arbitrary $\alpha$ and
$\beta$ using (\ref{5-2}) and the fusion rules
and found it in exact agreement
with our conjecture when setting $\alpha=1-\beta$. As a second,
independent test we set $\eta_x=1(0)$ corresponding to the bulk value
in the high density region (low density
region) and indeed obtained $G_c(x_1,x_2;4)=0$ for arbitrary $\alpha$
and $\beta$.

The choice $\alpha=1-\beta$ is not too restrictive as far as the
physics is concerned: since this curve runs across the phase diagram
it covers both the high density phase and the low density phase
and crosses the phase transition line at $\alpha=\beta=1/2$.
In what follows we study $G(x_1,x_2;t)$ in the
low density phase along the curve $\beta=1-\alpha>1/2$
and on the phase transition line at $\beta=1/2$.

In the low density phase we focus on the boundary region
with a non-uniform density profile.
For $\beta>1/2,\; \alpha=1-\beta$ the expression (\ref{3-4})
for the density profile for large $L$  gives $ \mbox{$\langle \,
\eta_{2x-1} \, \rangle$}=
 \mbox{$\langle \, \eta_{2x} \, \rangle$} =((1-\beta)/\beta)^{L+2-2x}$
 and therefore
\begin{equation}\label{5-8}
\beta \mbox{$\langle \, \eta_{x+3-2t+2k} \, \rangle$}+(1-\beta)
\mbox{$\langle \, \eta_{x+4-2t+2k} \, \rangle$} =
\left( \mbox{\small$\frac{1-\beta}{\beta}$}
 \right)^{L+1-x+2t-2k} \hspace*{2mm} .
\end{equation}
Inserting this into (\ref{5-7}) and introducing the incomplete
$\beta$-function
\begin{equation}\label{5-9}\begin{array}{rcl}
\displaystyle\sum_{k=0}^{t+y-2}
{\renewcommand{\arraystretch}{0.6}
\mbox{$\left(\begin{array}{@{}c@{}}{\scriptstyle 2t-2}\\
{\scriptstyle k}\end{array}\right)$}
\renewcommand{\arraystretch}{1}}
\beta^{2t-2-k}(1-\beta)^k & = &
I_{\beta}(t-y,t+y-1) \vspace*{4mm} \\
& = & \displaystyle 1 - I_{1-\beta}(t+y,t-y) -
{\renewcommand{\arraystretch}{0.6}
\mbox{$\left(\begin{array}{@{}c@{}}{\scriptstyle 2t-2}\\
{\scriptstyle t-y-1}\end{array}\right)$}
\renewcommand{\arraystretch}{1}}
\beta^{t-y}(1-\beta)^{t+y-1}
\end{array}\end{equation}
the correlation function (\ref{5-7}) can conveniently be rewritten
\begin{equation}\label{5-10}
G(x,x+2y;t) =
\left(\mbox{\small$\frac{1-\beta}{\beta}$}
\right)^{L-x} \left( I_{1-\beta}(t-y,t+y)
+ \left(\mbox{\small$\frac{\beta}{1-\beta}$}
 \right)^{2y-1} I_{1-\beta}(t+y,t-y)
 \right) \hspace*{2mm} .
\end{equation}

For large times $t$  (such that $|y|/t\ll 1$) the incomplete
$\beta$-function has the asymptotic form
\begin{equation}\label{5-11}
I_{1-\beta}(t+y,t-y) = \left\{ \begin{array}{lr} \displaystyle
(1-\frac{3}{4\xi_t}) \, \mbox{e}^{1/\xi_t} \,
\sqrt{\frac{\xi_t}{4\pi t}} \, \mbox{e}^{-(y/\xi_r+t/\xi_t)} \,
\mbox{e}^{-y^2/t} & \beta > \mbox{\small$\frac{1}{2}$} \vspace*{4mm} \\
\displaystyle P(\mbox{\small$\frac{y}{\sqrt{t}}$})-
\sqrt{\frac{t}{\pi \xi_t}}
\, \mbox{e}^{-y^2/t} & \beta = \mbox{\small$\frac{1}{2}$} +
(2\xi_t)^{-1/2}
 \end{array} \right.
\end{equation}
with
\begin{equation}\label{5-12}
\xi_t^{-1} = - \ln{(4\beta(1-\beta))}, \hspace*{6mm}
\xi_r^{-1} = - \ln{\mbox{\small$\frac{1-\beta}{\beta}$}}\hspace*{2mm} .
\end{equation}
and the probability integral $P(u)=1/\sqrt{2\pi} \int_{-\infty}^{u}
\exp{(-t^2/2)}dt$. In terms of $\xi_t$ the inequality $\beta>1/2$
in the upper expression of the r.h.s. of (\ref{5-11})
has to be understood as $1 \ll \xi_t
\raisebox{-1.0mm}{\mbox{$\stackrel{\textstyle <}{\sim}$}}
t$. In the lower
expression we assume $1 \ll t
\raisebox{-1.0mm}{\mbox{$\stackrel{\textstyle <}{\sim}$}}
\xi_t$.
Note that the two length scales $\xi_t$ and $\xi_r$ are not
independent quantities but related through $\xi_t^{-1} =
\ln \cosh^2(\xi_r^{-1}/2)$. As $\beta$ approaches $1/2$,
$\xi_t$ and $\xi_r$ diverge and are asymptotically related through
$\xi_t \approx 4 \xi_r^2$.

We define $\xi=\xi_r$ and $r=|y|=1/2|x_2-x_1|$ and insert (\ref{5-11})
into (\ref{5-10}). This gives the
scaling form of the time-dependent correlation function in the
scaling region of large $2\xi
\raisebox{-1.0mm}{\mbox{$\stackrel{\textstyle <}{\sim}$}}
t^{1/2}$
\begin{equation}\label{5-13}
G(x_1,x_2;t) = \mbox{e}^{-R/\xi}\,\mbox{e}^{-r/\xi}\,
\sqrt{\frac{4\xi^2}{\pi t}}\, \mbox{e}^{-t/(4\xi^2)}\,
\mbox{e}^{-r^2/t}
\end{equation}
where
\begin{equation}\label{5-13a}
R = \left\{ \begin{array}{ll} L+1-x_2  &
\mbox{if $2y=x_2-x_1>0$} \vspace*{4mm} \\
L+1-x_1  & \mbox{if $2y=x_2-x_1<0$} \end{array} \right.
\end{equation}
measures the distances of $x_2$ or $x_1$ from the boundary, depending
on the sign of $x_2-x_1$.
$G(x_1,x_2;t)$ is invariant under the scaling transformation
\begin{equation}\label{5-14}\begin{array}{ccc}
R \rightarrow \lambda R, \hspace*{4mm} &  r \rightarrow \lambda r,
\hspace*{4mm} &
\xi \rightarrow \lambda \xi \vspace*{4mm} \\
 & t \rightarrow \lambda^2 t & \hspace*{2mm} .
\end{array}\end{equation}
This is of the form corresponding to dynamical scaling  with a
dynamic critical exponent $z=2$ and critical exponent $x=0$.

If $\xi$ increases beyond the crossover length scale $t^{1/2}$
the correlation function changes its form.
At the critical point $\beta=1/2$ we have $ \mbox{$\langle \, \eta_x \,
\rangle$}=x/L$
and up to corrections of order $1/L$ the exact expression (\ref{5-7})
for the correlation function gives
\begin{equation}\label{5-15}\begin{array}{rcl}
G(x,x+2y;t) & = & \displaystyle \frac{x+2y}{L} - \frac{2t+2y}{L}
I_{\mbox{\small$\frac{1}{2}$}}(t-y,t+y-1)+\vspace*{4mm} \\
 & & \displaystyle \frac{2}{L} \left(\frac{1}{2}\right)^{2t-2}
 \sum_{k=0}^{t+y-2} k
{\renewcommand{\arraystretch}{0.6}
\mbox{$\left(\begin{array}{@{}c@{}}{\scriptstyle 2t-2}\\
{\scriptstyle k}\end{array}\right)$}
\renewcommand{\arraystretch}{1}}
 \hspace*{2mm} .
\end{array}\end{equation}
Using
\begin{equation}\label{5-16}\begin{array}{rcl}
\displaystyle 2\sum_{k=0}^{t+y-2} k
{\renewcommand{\arraystretch}{0.6}
\mbox{$\left(\begin{array}{@{}c@{}}{\scriptstyle 2t-2}\\
{\scriptstyle k}\end{array}\right)$}
\renewcommand{\arraystretch}{1}}
\beta^{2t-2-k}
(1-\beta)^k & = &
2(1-\beta)(2t-2)I_{\beta}(t-y,t+y-1) - \vspace*{4mm} \\
 & & 2(t+y-1)
{\renewcommand{\arraystretch}{0.6}
\mbox{$\left(\begin{array}{@{}c@{}}{\scriptstyle 2t-2}\\
{\scriptstyle t-y-1}\end{array}\right)$}
\renewcommand{\arraystretch}{1}}
 \beta^{t-y}(1-\beta)^{t+y-1}
\end{array}\end{equation}
and the expansion (\ref{5-11}) of
$I_{\mbox{\small$\frac{1}{2}$}}(t+y,t-y)$ we obtain with
$2y=x_2-x_1$ and $u=y/\sqrt{t}$
\begin{equation}\label{5-17}\begin{array}{rcl}
G(x_1,x_2;t) & = & \displaystyle
\frac{x_1+x_2}{2L} - \frac{t}{L} \left( \frac{1}{2} \right)^{2t-2}
{\renewcommand{\arraystretch}{0.6}
\mbox{$\left(\begin{array}{@{}c@{}}{\scriptstyle 2t-2}\\
{\scriptstyle t-y-1}\end{array}\right)$}
\renewcommand{\arraystretch}{1}}
- \frac{y}{L}\left(
1-2I_{\mbox{\small$\frac{1}{2}$}}(t-y,t+y)\right)
\vspace*{4mm} \\
 & \approx & \displaystyle\frac{x_1+x_2}{2L} -
 \frac{\sqrt{t}}{\sqrt{\pi}L}
 \left(\mbox{e}^{-u^2} + \sqrt{\pi} u (1 - 2P(u))\right).
 \hspace*{2mm} .
\end{array}\end{equation}
For finite distances $y,t$ one has (up to corrections of order $L^{-1}$
which we neglect) $x_1=x_2=x$ and the correlation function at the
critical point is space
and time independent with an amplitude $x/L$ depending on the
relative position of $x$ in the bulk.
For large times, $t\propto L$ and $y/t\ll 1$, the correlation function
gets an contribution order $L^{-1/2}$. $G(x_1,x_2;t)$
is invariant under the scale transformations (\ref{5-14})
with the length scale $\xi$ replaced by the size of the system $L$.

This result has a simple interpretation in terms of the constituent
profiles discussed in the preceding section. For simplicity we
consider
$y=0$. The operator $\eta_{x}T^t\eta_{x}$ gives 1 if $x$ is in
a region of high density both at times $t_0$ and $t_0+t$. The
probability that $x$ is in the high density region  at time $t_0$
is $x/L$. This accounts for the constant $x/L$ in (\ref{5-17})
If we
assume that the domain wall separating the low density region
from the high density region performs a random walk of two lattice
units per time step starting
from its position $x_0$ at time $t_0$ then the resulting expectation
value  \mbox{$\langle \, \eta_xT^t\eta_x \, \rangle$}
will indeed be of the form (\ref{5-17}).

\section{Summary and comparison with other exclusion models}

Let us summarize our main results. We obtained recursion rules
(\ref{2-6}) which allow the construction of the steady state
of the model
defined by (\ref{1-1}) and (\ref{1-2}). Moreover we found for
arbitrary values $\alpha$ and $\beta$ of the injection and absorption
rates exact steady state
expressions for the density-profile (\ref{2-10}), the current
(\ref{2-11}), the equal-time $n$-point density correlation function
(\ref{4-3}) and the time dependent two-point function eqs.
(\ref{5-3}) - (\ref{5-7}).\footnote{The expression (\ref{5-7})
for the time-dependent two-point function inside the light cone
was found only for $\alpha=1-\beta$.}

We made the following observations:\\
(a) The phase diagram (fig.~1) shows two phases. In the low density
phase ($\alpha<\beta$) the average density is $\rho=\alpha/2$ (in the
limit $L\rightarrow \infty$) and in the high density phase
one has $\rho=1-\beta/2$.
On the phase transition line $\alpha=\beta$ the average density
is $\rho=1/2$.\\
(b) There is a shift in the average densities between the even
sublattice
and the odd sublattice (see eqs. (\ref{3-5}) - (\ref{3-7})). This shift
is the current $j$. In terms of the total average density $\rho$
one finds $j=2\rho$ in the low density phase and $j=2(1-\rho)$ in the
high density phase. On the phase transition line ($\rho=1/2$)
the current is $j=\alpha$ (i.e. $j\neq 1=2\rho=2(1-\rho)$).\\
(c) In the low density phase the density profile decays on both
sublattices to its respective bulk value exponentially with
increasing distance from the boundary while in the high density phase
it increases exponentially to its respective bulk value with
increasing
distance from the origin. At the phase transition line $\alpha=\beta$
the length scale $\xi$ (\ref{3-4}) associated with the exponential
shape of the profile diverges and the profile increases linearly
on both sublattices.
The average density and the first derivatives of the current w.r.t.
$\alpha$ and $\beta$
have a discontinuity at the phase transition
line (in the thermodynamic limit $L\rightarrow \infty$).\\
(d) The decay length $\xi$ is identical with the correlation length
of the connected two-point density correlation function. Outside the
light cone this correlator is of the scaling form $G(x_1,x_2;t=0) =
A r^{\kappa} \exp{(-r/\xi)}$ with exponent $\kappa=0$.
The amplitude $A$ is space-dependent as  a result of breaking of
translational invariance (\ref{4-5}) and (\ref{4-6}).
(e) The time-dependent two-point correlation function $G(x_1,x_2;t)$
inside the light cone near the critical line (\ref{5-13})
has a form compatible with dynamical scaling with dynamic critical
exponent $z=2$ and scaling dimension $x=0$ (see the transformations
(\ref{5-14})). It contains the exponential $\exp{(-r^2/t)}$
characteristic for local dynamical scaling, but the amplitude is
again space-dependent due to breaking of translational invariance.
The correlation function changes its form when the correlation length
increases beyond a crossover length scale of order $t^{1/2}$.
On the critical line it becomes a space-dependent constant up
to corrections of order $L^{-1}$ if $t$ is large but finite,
and up to corrections of order $L^{-1/2}$ for times of order $L$.\\
(f) From an analysis of the two-point correlation function we found
that the density profile can be considered as a superposition
of step-function type profiles with average density
$\rho_1^{(even)}=\alpha$ and $\rho_1^{(odd)}=0$ up to some
point $x_0$ and average density
$\rho_2^{(even)}=1$ and $\rho_2^{(odd)}=1-\beta$ beyond this
point up to the boundary. These densities are the sublattice
densities in the low density phase and high density phase
respectively. The probability of finding the
``domain wall'' at site $x_0$ separating the two regions of high
density and low density decreases exponentially
with increasing distance from the boundary (origin) in the low density
(high density) phase. On the phase transition line this probability
is space independent and the domain wall can be found everywhere
with the same probability. Studying the time-dependent correlation
function suggests that it performs a random walk around its
position $x_0$ at time $t_0$.

We conclude our discussion of the deterministic p/o-model with a
brief comparison with other exclusion models and some conjectures
for probabilistic exclusion models with parallel dynamics.

In the deterministic p/p-model with a defect we found a phase
diagram showing a
low density phase, a coexistence phase where a low density density
region coexists with a high density region and a high density phase.
The relation between the current and the sublattice densities and
the total density discussed in (b) for the low and high density phases
is identical with that in the p/p-model in the respective phases.
This is a consequence of the bulk dynamics which are identical
in both cases and could have been guessed.

It is interesting to observe
that there is also a correspondence between the coexistence phase of
the p/p-model and the phase transition line here: In the p/p-model
with defect strength $1-q$ the quantity $q$ is the hopping probability
at a single link of the ring, say between sites $L$ and 1 and
therefore corresponds to an absorption of particles
at site $L$ and injection of particles at site 1 with rate $q$.
This injection and absorption is correlated because of the
particle number conservation.
In the coexistence phase the current is density-independent and one
has $j=q$. In the p/o-model on the phase transition line discussed
here particles are also injected and absorbed with the same rate
$\alpha$, but uncorrelated. The current is density-independent,
$j=\alpha$ as in the p/p-model. The profile is build by constituent
profiles with a region of low density up to some $x_0$ and a high
density region beyond that point. $x_0$ can be anywhere with same
probability. This suggests that also the profile in the coexistence
phase of the p/p-model with defect is build by such constituent
profiles, with the distinction that there $x_0$ cannot be anywhere
(because of particle number conservation) but the probability $p(x_0)$
of finding the domain wall is centered around some point $R_0$.
We believe that similar phases occur also in probabilistic p/p-models
with a defect and in probabilistic p/o-models. In such models
particles do not always move if the neighbouring site was empty as here
but can stay with some non-zero probability even in the bulk.
For mixed models (defect and some uncorrelated injection and
absorption) we expect correspondingly a softening of the distribution
$p(x_0)$.

All the features summarized under (c)
(except the implied anisotropy between the even and odd sublattices)
are in common with the phase transition from
the low density phase ${\rm A}_{\rm I}$ to the high density phase
${\rm B}_{\rm I}$ in the s/o-model \cite{3}. This suggests that
also the correlation functions are qualitatively the
same (in the thermodynamic limit $L\rightarrow \infty$)
near the phase transition
line. In the s/o-model
however, the transition line $\alpha=\beta$ extends only up to
$\alpha=\beta=1/2$ as opposed to $\alpha=\beta=1$ here.
Correspondingly, in our model there
is no maximal current phase with a power behaviour of the
density profile and no phases corresponding to the phases
${\rm A}_{\rm II}$ or ${\rm B}_{\rm II}$ (for $\alpha>1/2$ or
$\beta>1/2$) of \cite{3} where the shape of the density profile
is determined by a product of a power law behaviour with an
exponential decay.

We believe that
phase transitions to such phases cannot occur in our model because
of the deterministic nature of the dynamics. In \cite{3} we argued
that these phases result from an ``overfeeding'' of the system
with particles: The system reaches its maximal transport capacity
at density 1/2. In order to obtain average density 1/2 at the
origin particles have to be injected with rate 1/2.
If particles are injected at a higher rate they
block each other rather than moving into the bulk
and cause a phase transition. Here this cannot happen.
The system reaches its maximal transport capacity
also at average density 1/2  but here this corresponds to a
completely filled even sublattice and an empty odd sublattice. An
average density of 1 on the even sublattice at the origin can only
occur if particles are injected at rate 1. Thus there can be no
``overfeeding''. The deterministic hopping rules imply that particles
injected at the origin move away with velocity of light,
therefore there can be no mutual blockage near the origin.
(Similar arguments can be used for a discussion of the dependence of
the phase transitions on the absorption rate $\beta$ by exchanging
particles with holes and studying the injection of holes at the
boundary).

This discussion naturally leads to the conjecture that the
probabilistic p/o-model (which has not yet been studied)
will have a phase diagram similar to that
of the s/o-model with a phase transition line at $\alpha=\beta$
up to some value $\alpha_0$ where the profile is constant on
both sublattices and additional phase transition lines at the
values of $\alpha$ and $\beta$ corresponding to an overfeeding
of particles at the origin and holes at the boundary respectively.

\subsection*{Acknowledgments}

It is a  pleasure to thank E. Domany and D. Mukamel for interesting
dicussions. Financial support by the Deutsche Forschungsgemeinschaft
(DFG) is gratefully acknowledged.

\appendix
\section{Mapping to a two-dimensional vertex model}

Following the idea of ref. \cite{KDN} we want to show how the
exclusion process defined by relations (\ref{1-1}) and (\ref{1-2})
is related to a two-dimensional vertex model.
The discussion is partly similarly to that in \cite{1} which we
repeat for the convenience of the reader unfamiliar
with this mapping. The mapping of the boundary conditions to vertices
in the vertex model is different from \cite{1}.

Consider a 4-vertex model on a diagonal square lattice defined
as follows: Place an up- or down-pointing arrow on each link of
the lattice and assign a non-zero Boltzmann weight to each of
the vertices shown in figure~1. (All other configurations of
arrows around an intersection of two lines, i.e., all other vertices,
are forbidden in the bulk.) The partition function is the sum of
the products
of Boltzmann weights of a lattice configuration taken over all
allowed configurations.
\begin{center}
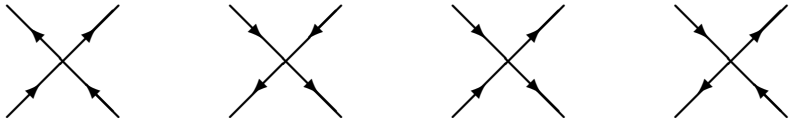
\begin{figure}[h]
\setlength{\unitlength}{3.7mm}
\begin{center}
\begin{picture}(40,10)
\thicklines
% vertex 1
\put (4,3){\line(1,1){4.}}
\put (4,7){\line(1,-1){4.}}
\put (5.2,4.2){\vector(1,1){0}}
\put (7.2,6.2){\vector(1,1){0}}
\put (4.8,6.2){\vector(-1,1){0}}
\put (6.8,4.2){\vector(-1,1){0}}
% vertex 2
\put (12,3){\line(1,1){4.}}
\put (12,7){\line(1,-1){4.}}
\put (12.8,3.8){\vector(-1,-1){0}}
\put (14.8,5.8){\vector(-1,-1){0}}
\put (15.2,3.8){\vector(1,-1){0}}
\put (13.2,5.8){\vector(1,-1){0}}
% vertex 4
\put (20,3){\line(1,1){4.}}
\put (20,7){\line(1,-1){4.}}
\put (21.2,4.2){\vector(1,1){0}}
\put (23.2,6.2){\vector(1,1){0}}
\put (23.2,3.8){\vector(1,-1){0}}
\put (21.2,5.8){\vector(1,-1){0}}
%% vertex 5
\put (28,3){\line(1,1){4.}}
\put (28,7){\line(1,-1){4.}}
\put (28.8,3.8){\vector(-1,-1){0}}
\put (31.2,6.2){\vector(1,1){0}}
\put (29.2,5.8){\vector(1,-1){0}}
\put (30.8,4.2){\vector(-1,1){0}}
\put (5.8,0){$a_1$}
\put (13.8,0){$a_2$}
\put (21.8,0){$b_2$}
\put (29.8,0){$c_2$}
\end{picture}
\end{center}
\caption{\protect\small
Allowed bulk vertex configurations in the four-vertex model.
Up-pointing arrows correspond to particles, down-pointing
arrows represent vacant sites. In the dynamical interpretation
of the model the Boltzmann weights give
the transition probability of the state represented by the
pair of arrows below the vertex to that above the vertex.}
\end{figure}
\end{center}

In the transfer matrix formalism up- and down-pointing arrows
in each row of a
diagonal square lattice built by $M$ of these vertices represent the
state of the system at some given time $t$.
Corresponding to the $M$ vertices there are $L=2M$ sites in each row
represented by the links of the diagonal lattice.
The configuration of arrows in the next row above (represented by the
upper arrows of the same vertices) then corresponds to the state of
the system at an intermediate time $t'=t+1/2$, and the
configuration after a full time step $t''=t+1$ corresponds to the
arrangement of arrows two rows above.
Therefore each vertex represents a local transition
from the state given by the lower two arrows of a vertex representing
the configuration on sites $j$ and $j+1$ at time $t$
to the state defined by the upper two arrows representing the
configuration at sites $j$ and $j+1$ at time $t+1/2$.
The correspondence of the vertex language to the particle picture
used in the introduction can be understood by considering
 up-pointing arrows as
particles occupying the respective sites of the chain
while down-pointing arrows represent vacant sites, i.e., holes.

The diagonal-to-diagonal transfer matrix $T$
acting on a chain of $L$ sites ($L$ even)
of the vertex model with vertex weights
$a_1, \dots  ,c_1$ as shown in fig.~2 is then defined by \cite{1,Vega}
\begin{equation}\label{A-1}
T = \prod_{j=1}^{L/2} T_{2j-1} \cdot \prod_{j=1}^{L/2} T_{2j} =
    T^{\rm odd} \, T^{\rm even} \hspace*{2mm} .
\end{equation}
The matrices $T_j$ act nontrivially on sites $j$ and $j+1$ in the
chain, on all other sites they act as unit operator.
All matrices $T_j$ and $T_{j'}$ with $|j-j' | \neq 1$ commute.
(The difference $j-j'$ is understood to be mod $L$).
For an explicit representation of the transfer matrix
we choose a spin-1/2
tensor basis where the Pauli-matrix $\sigma_j^z$ acting on site
$j$ of the chain is diagonal and spin down at site $j$ represents
a particle (up-pointing arrow) and spin up a hole (down-pointing
arrow). In this basis
$\tau_j = \mbox{\small$\frac{1}{2}$} (1 - \sigma_j^{z})$ is
the projection operator
on particles on site $j$,
$\sigma_j = \mbox{\small$\frac{1}{2}$} (1 + \sigma_j^{z})$ is
the projector on holes and
$s_j^{\pm} = \mbox{\small$\frac{1}{2}$} ( \sigma_j^x \pm i
\sigma_j^y )$
($\sigma^{x,y,z}$ being the Pauli matrices) create ($s_j^{-}$)
and annihilate ($s_j^{+}$) particles respectively.

The bulk dynamics of our model is encoded in the transfer matrix
by choosing the vertex weights as
\begin{equation}\label{A-2}
a_1 = a_2 = b_2 = c_1 = 1
\end{equation}
In the bulk this leads to
\begin{equation}\label{A-3}
T_j = 1 + s_j^{+} s_{j+1}^{-} - \tau_j \sigma_{j+1} =
\left( \begin{array}{cccc}
       1\hspace*{2mm} & 0\hspace*{2mm} & 0\hspace*{2mm} & 0 \\
       0 & 1 & 1 & 0 \\
       0 & 0 & 0 & 0 \\
       0 & 0 & 0 & 1 \end{array} \right)_{j,j+1} \hspace*{2mm} .
\end{equation}
In the particle language
the matrices $T_j$ describe the local transition probabilities
of particles moving from site $j$ to site $j+1$ represented
by the corresponding vertices. If sites $j$ and $j+1$ are both
empty or occupied, they remain as they are under the action of
$T_j$. The same holds for a hole on site $j$ and a particle
on site $j+1$, corresponding to the diagonal elements of $T_j$,
representing vertices $a_1$, $a_2$ and $c_1$.
If there is a particle on site $j$ and a
hole on site $j+1$, the particle will move with probability
one to site $j+1$. This accounts for vertex $b_2$.

As discussed in the introduction we
assume  open boundary conditions with injection of particles
on site 1 and absorption
of particles on site $L$. This allows for the additional
vertices shown in fig.~3 together with vertex weights corresponding
to the respective probabilities of creating and annihilating particles.
\begin{center}
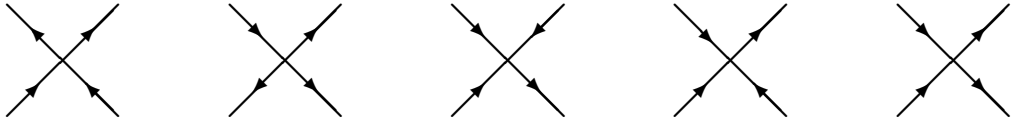
\begin{figure}[h]
\setlength{\unitlength}{3.7mm}
\begin{center}
\begin{picture}(40,10)
\thicklines
% vertex 1
\put (0,3){\line(1,1){4.}}
\put (0,7){\line(1,-1){4.}}
\put (1.2,4.2){\vector(1,1){0}}
\put (3.2,6.2){\vector(1,1){0}}
\put (0.8,6.2){\vector(-1,1){0}}
\put (2.8,4.2){\vector(-1,1){0}}
% vertex 2
\put (8,3){\line(1,1){4.}}
\put (8,7){\line(1,-1){4.}}
\put (8.8,3.8){\vector(-1,-1){0}}
\put (11.2,6.2){\vector(1,1){0}}
\put (11.2,3.8){\vector(1,-1){0}}
\put (9.2,5.8){\vector(1,-1){0}}
%% vertex 3
\put (16,3){\line(1,1){4.}}
\put (16,7){\line(1,-1){4.}}
\put (17.2,4.2){\vector(1,1){0}}
\put (18.8,5.8){\vector(-1,-1){0}}
\put (17.2,5.8){\vector(1,-1){0}}
\put (19.2,3.8){\vector(1,-1){0}}
%% vertex 4
\put (24,3){\line(1,1){4.}}
\put (24,7){\line(1,-1){4.}}
\put (25.2,4.2){\vector(1,1){0}}
\put (27.2,6.2){\vector(1,1){0}}
\put (25.4,5.6){\vector(1,-1){0}}
\put (26.8,4.2){\vector(-1,1){0}}
% vertex 5
\put (32,3){\line(1,1){4.}}
\put (32,7){\line(1,-1){4.}}
\put (33.2,4.2){\vector(1,1){0}}
\put (35.2,6.2){\vector(1,1){0}}
\put (35.2,3.8){\vector(1,-1){0}}
\put (33.2,5.8){\vector(1,-1){0}}
\put (1.8,0){$\beta$}
\put (9.8,0){$\alpha$}
\put (15.8,0){$\beta(1-\alpha)$}
\put (23.8,0){$\alpha(1-\beta)$}
\put (31.8,0){$(1-\alpha)(1-\beta)$}
\end{picture}
\end{center}
\caption{\protect\small
Additional vertex configurations allowed at the boundary
and their Boltzmann weights.
The left arrows of these vertices describe the particle configuration
at the boundary site $L$ of the system while the right arrows
define the particle configurations at the origin (site 1).}
\end{figure}
\end{center}

In a two-dimensional lattice (fig.~4) we consider the half-vertices
at the left boundary as the right arms of the vertices shown in
fig.~2 and fig.~3
and the half-vertices at the right boundary as their left arms.
Thus the left arrows define the particle configuration on site
$L$ and the right arrows are considered as site 1.
Vertices $a_1$, $a_2$ and $b_2$ have a different weight at the
boundary: $a_1'=1-\beta$, $a_2'=1-\alpha$, $b_2'=\alpha\beta$.
Note that vertex $b_2$ at the boundary describes simultaneous
absorption of a particle at site $L$ and creation of a particle
at site 1.

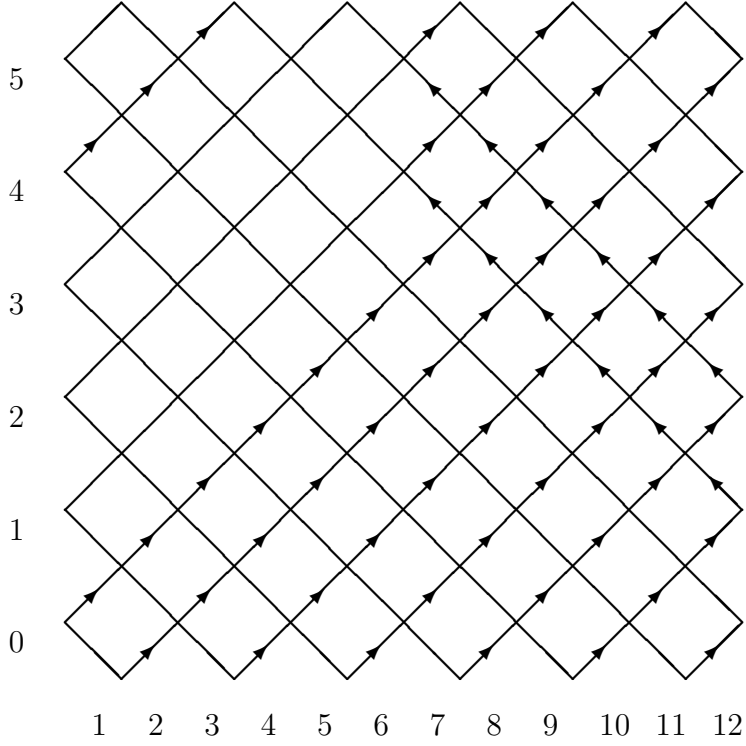
\begin{figure}
\setlength{\unitlength}{2.5mm}
\begin{center}
\begin{picture}(42,42)
\thicklines
\put (6,3){\line(1,1){33.}}
\put (12,3){\line(1,1){27.}}
\put (18,3){\line(1,1){21.}}
\put (24,3){\line(1,1){15.}}
\put (30,3){\line(1,1){ 9.}}
\put (36,3){\line(1,1){ 3.}}
\put (3,6){\line(1,1){33.}}
\put (3,12){\line(1,1){27.}}
\put (3,18){\line(1,1){21.}}
\put (3,24){\line(1,1){15.}}
\put (3,30){\line(1,1){ 9.}}
\put (3,36){\line(1,1){ 3.}}
\put (3,6){\line(1,-1){3.}}
\put (3,12){\line(1,-1){ 9.}}
\put (3,18){\line(1,-1){15.}}
\put (3,24){\line(1,-1){21.}}
\put (3,30){\line(1,-1){27.}}
\put (3,36){\line(1,-1){33.}}
\put (6,39){\line(1,-1){33.}}
\put (12,39){\line(1,-1){27.}}
\put (18,39){\line(1,-1){21.}}
\put (24,39){\line(1,-1){15.}}
\put (30,39){\line(1,-1){ 9.}}
\put (36,39){\line(1,-1){ 3.}}
\put ( 4.8,31.8){\vector(1,1){0}}
\put ( 7.8,34.8){\vector(1,1){0}}
\put (10.8,37.8){\vector(1,1){0}}
\put (22.8,31.8){\vector(1,1){0}}
\put (25.8,34.8){\vector(1,1){0}}
\put (28.8,37.8){\vector(1,1){0}}
\put (22.8,37.8){\vector(1,1){0}}
\put ( 4.8,7.8){\vector(1,1){0}}
\put ( 7.8,10.8){\vector(1,1){0}}
\put (10.8,13.8){\vector(1,1){0}}
\put (13.8,16.8){\vector(1,1){0}}
\put (16.8,19.8){\vector(1,1){0}}
\put (19.8,22.8){\vector(1,1){0}}
\put (22.8,25.8){\vector(1,1){0}}
\put (25.8,28.8){\vector(1,1){0}}
\put (28.8,31.8){\vector(1,1){0}}
\put (31.8,34.8){\vector(1,1){0}}
\put (34.8,37.8){\vector(1,1){0}}
\put (7.8,4.8){\vector(1,1){0}}
\put (10.8,7.8){\vector(1,1){0}}
\put (13.8,10.8){\vector(1,1){0}}
\put (16.8,13.8){\vector(1,1){0}}
\put (19.8,16.8){\vector(1,1){0}}
\put (22.8,19.8){\vector(1,1){0}}
\put (25.8,22.8){\vector(1,1){0}}
\put (28.8,25.8){\vector(1,1){0}}
\put (31.8,28.8){\vector(1,1){0}}
\put (34.8,31.8){\vector(1,1){0}}
\put (37.8,34.8){\vector(1,1){0}}
\put (13.8, 4.8){\vector(1,1){0}}
\put (16.8, 7.8){\vector(1,1){0}}
\put (19.8,10.8){\vector(1,1){0}}
\put (22.8,13.8){\vector(1,1){0}}
\put (25.8,16.8){\vector(1,1){0}}
\put (28.8,19.8){\vector(1,1){0}}
\put (31.8,22.8){\vector(1,1){0}}
\put (34.8,25.8){\vector(1,1){0}}
\put (37.8,28.8){\vector(1,1){0}}
\put (19.8, 4.8){\vector(1,1){0}}
\put (22.8, 7.8){\vector(1,1){0}}
\put (25.8,10.8){\vector(1,1){0}}
\put (28.8,13.8){\vector(1,1){0}}
\put (31.8,16.8){\vector(1,1){0}}
\put (34.8,19.8){\vector(1,1){0}}
\put (37.8,22.8){\vector(1,1){0}}
\put (25.8, 4.8){\vector(1,1){0}}
\put (28.8, 7.8){\vector(1,1){0}}
\put (31.8,10.8){\vector(1,1){0}}
\put (34.8,13.8){\vector(1,1){0}}
\put (37.8,16.8){\vector(1,1){0}}
\put (31.8, 4.8){\vector(1,1){0}}
\put (34.8, 7.8){\vector(1,1){0}}
\put (37.8,10.8){\vector(1,1){0}}
\put (37.8, 4.8){\vector(1,1){0}}
\put (22.2,28.8){\vector(-1,1){0}}
\put (25.2,25.8){\vector(-1,1){0}}
\put (28.2,22.8){\vector(-1,1){0}}
\put (31.2,19.8){\vector(-1,1){0}}
\put (34.2,16.8){\vector(-1,1){0}}
\put (37.2,13.8){\vector(-1,1){0}}
\put (22.2,34.8){\vector(-1,1){0}}
\put (25.2,31.8){\vector(-1,1){0}}
\put (28.2,28.8){\vector(-1,1){0}}
\put (31.2,25.8){\vector(-1,1){0}}
\put (34.2,22.8){\vector(-1,1){0}}
\put (37.2,19.8){\vector(-1,1){0}}
\put (4.4,0){1}
\put (7.4,0){2}
\put (10.4,0){3}
\put (13.4,0){4}
\put (16.4,0){5}
\put (19.4,0){6}
\put (22.4,0){7}
\put (25.4,0){8}
\put (28.4,0){9}
\put (31.4,0){10}
\put (34.4,0){11}
\put (37.4,0){12}
\put (0,4.4){0}
\put (0,10.4){1}
\put (0,16.4){2}
\put (0,22.4){3}
\put (0,28.4){4}
\put (0,34.4){5}
\end{picture}
\end{center}
\caption{\protect\small
Configuration of particles (up-pointing arrows) on a lattice
of length $L=12$ in space (horizontal) direction $M=2t=12$ between
times $t=0$ and $t=5+1/2$ (vertical direction). Down-pointing arrows
denoting vacant sites have been omitted from the drawing. At time
$t=0$ the even sublattice is filled and the odd sublattice empty.
Particles are injected at site 1 after times $t=0$ and $t=4$. At the
boundary (site 12) particles get stuck at times $t=1$ and $t=2$
and are absorbed at times $t=0,3,4,5$.}
\end{figure}

With this convention $T_L(\alpha,\beta)$ acting on sites $L$ and 1
corresponding to the vertex weights shown in
fig.~3 is given by
\begin{equation}\label{A-4}\begin{array}{rcl}
T_L(\alpha,\beta)  & = &
1 + \alpha (s_1^{-} - \sigma_1) + \beta (s_L^{+} - \tau_L)
+ \alpha\beta (s_L^{+} - \tau_L) (s_1^{-} - \sigma_1) \vspace*{4mm} \\
 & = & \displaystyle
\left( \begin{array}{cccc}
1-\alpha & 0 & \beta(1-\alpha)     & 0     \\
\alpha   & 1 & \alpha \beta        & \beta \\
0        & 0 & (1-\alpha)(1-\beta) & 0     \\
0        & 0 & \alpha(1-\beta)     & 1-\beta \end{array}\right)_{L,1}
\hspace*{2mm} .
\end{array}\end{equation}

The transfer matrix $T=T(\alpha,\beta)$
acts parallel first on all even-odd pairs of sites $(2j,2j+1)$
including the boundary pair ($L$,1),
then on all odd-even pairs. Thus in the first half time step
$T^{\rm even}$ shifts particles from the even
sublattice to the odd sublattice (so far it was not
occupied) and then, in the second half step,
$T^{\rm odd}$ moves particles from
the odd sublattice to the even sublattice again. As a result,
we expect an asymmetry in the average occupation of the even and
odd sublattice which is related to the particle current.
In a model with transfer matrix $\tilde{T} =
T^{\rm odd} T^{\rm even}$ the asymmetry will be reversed, but
there will be no essential difference in the
physical properties of these two systems.

A possible configuration of particles in a 12x12 lattice is shown
fig.~4. Note that the presence of particles at site $x=11$ and times
$t=2,3$ imply the existence of particles on the left
edge of their light cones as long as they move in a region where
the even sublattice is fully occupied, i.e. they move with velocity
of light (two lattice units per time step) to the left.
A particle on an even lattice site at some (integer) time $t$ always
implies the existence of
a particle on the right edge of its light cone up to the
boundary.

The model has a particle hole symmetry.
We denote by $| x_1, x_2, \dots, x_N \rangle =
s_{x_1}^{-} s_{x_2}^{-} \dots s_{x_N}^{-} | \hspace*{2mm} \rangle$ the
$N$-particle state with particles on sites $x_1,\dots,x_N$
($| \hspace*{2mm} \rangle$ is the state with all spins up corresponding
to no particle).
The parity operator $P$ reflects particles with respect to the
center of the chain located between sites $x=L/2$ and $x=L/2+1$ and
the charge conjugation operator $C = \prod_{j=1}^{L} \sigma_{j}^{x}$
interchanges particles and holes and therefore turns a
$N$-particle state into a state with $L-N$ particles.
One finds
\begin{equation}
\label{A-5}
(CP)\,T(\alpha,\beta) \,(CP)= T(\beta,\alpha) \hspace*{2mm} .
\end{equation}

In the bulk the particle current is conserved and can be
obtained from the commutators of $\tau_{2x}$ and $\tau_{2x-1}$
with $T$.
These relations play a crucial role in the construction of
the steady state
and the computation of the time-dependent correlation function.
Defining the current operators $j_{2x}^{\rm even}$
and $j_{2x-1}^{\rm odd}$ by
\begin{equation}\label{A-6}
\begin{array}{rcll}
j_{2x}^{\rm even} & = & \tau_{2x}\sigma_{2x+1} \hspace*{6mm} &
(1\leq x \leq L/2-1)\vspace*{4mm}\\
j_{2x-1}^{\rm odd} & = &
(1-\sigma_{2x-2}\sigma_{2x-1})
(1-\tau_{2x}\tau_{2x+1}) & (2 \leq x \leq  L/2-1)
\end{array}\end{equation}
a straightforward calculation yields ($x\neq L/2$):
\begin{equation}\label{A-7}
\begin{array}{lcl}
\left[ T,\tau_{2x-1} \right] & = & \displaystyle
T \left( \tau_{2x-1} - (1 - \sigma_{2x-2}\sigma_{2x-1})
                      \tau_{2x}\tau_{2x+1} \right) \vspace*{4mm} \\
 & = & j_{2x-1}^{\rm odd} - j_{2x-2}^{\rm even} \vspace*{4mm} \\
\left[ T,\tau_{2x} \right] & = & \displaystyle
T\left( \sigma_{2x-2}\sigma_{2x-1}(1-\tau_{2x}\tau_{2x+1})-\sigma_{2x}
\right) \vspace*{4mm} \\
 & = & j_{2x}^{\rm even} - j_{2x-1}^{\rm odd} \hspace*{2mm} .
\end{array}\end{equation}

Current conservation implies that the expectation values of
the current operators
$j_{2x}^{\rm even}$ and $j_{2x-1}^{\rm odd}$
do not depend on $x$,
$\langle j_{2x}^{\rm even}\rangle = \langle j_{2x-1}^{\rm odd} \rangle
= const = j$.

Note that the cases $\alpha,\beta=0,1$ are trivial. If $\alpha=0$
no particles are injected and the steady state is
\mbox{$| \,  \, \rangle$}. If $\alpha=1$
then in each time step a particle is injected and therefore the even
sublattice fully occupied. Particles on the odd sublattice are
randomly distributed with average density $1-\beta$. As discussed
above they move with velocity of light everywhere. Therefore the
connected time-dependent two-point function on the odd sublattice
is 0 except on the left edge of the (forward) light cone.

\newpage
\bibliographystyle{unsrt}

\end{document}